# Impact of Three-Point Rule Change on Competitive Balance in Football: A Synthetic Control Method Approach

Ajay Sharma

Indian Institute of Management Indore

J-206, Academic Block, Prabandh Shikhar,

Rau-Pithampur Road, Indore (MP)-India-453556

e-mail: ajays@iimidr.ac.in

**Correspondence Address:**

**Ajay Sharma**, J-206, Academic Block, Indian Institute of Management Indore, Prabandh Shikhar, Rau-Pithampur Road, Indore, MP (India) - 453556. Ph: +91-7312439622. E-mail: ajays@iimidr.ac.in; ajaysharma87@gmail.com.

**Data availability statement**

The data used in this study is not publicly available. However, this data can be obtained from the authors upon request.

**Disclosure statement**

The authors reported no potential conflict of interest.

**Funding**

No funding was received by the authors for this research.

# Impact of Three-Point Rule Change on Competitive Balance in Football: A Synthetic Control Method Approach

Abstract

Governing authorities in sports often make changes to rules and regulations to increase competitiveness. One such change was made by the English Football Association in 1981 when it changed the rule for awarding points in the domestic league from two points for a win to three points. This study aims to measure this rule change's impact on the domestic league's competitive balance using a quasi-experimental estimation design of a synthetic control method. The three-point rule change led to an increase in competitive balance in the English League. Further, we show no significant change in the number of goals scored per match.

## 1. Introduction

In any sport, whether team or individual, increasing competitiveness is the core objective of any tournament. The governing or controlling authorities in a sport always seek ways to increase the participants' competitiveness (Sanderson, 2002; Lenten & Winchester, 2015; Percy, 2015; Parinduri et al., 2019; Butler et al., 2020). Across various sports, the authorities use various measures, such as changes in rules and laws, through which certainty in the outcome of the games played can be decreased, and in turn, the sport can be made more attractive for the spectators and fans. Some of these interventions include creating equality of opportunities for various teams through restrictions on resources that can be used in the competition, for example salary restriction (Fort & Lee, 2007; Booth, 2005; Annala & Winfree, 2011), draft process (Taylor & Trogdon, 2002; Frost et al., 2012), financial fair play in football (Peeters & Szymanski, 2014); creating or changing rules affecting the nature of play for example, offside rule in football, wide and no ball rule in cricket; changing the nature of tournament formats, such as round-robin format, home and away football matches (Page & Page, 2007), age rule (Vaeyens et al., 2005), false start disqualification in sprint (Brosnan et al., 2017), body checking rule in ice hockey (Hancock et al., 2011). Another mechanism is incentivizing the teams and players to exert more effort by awarding additional points or changing the symmetry of points won and lost in a drawn-versus-win situation.

In this study, we measure the impact of change in the point awarding rule on the competitive balance among teams in football tournaments. In 1981, the English Football Association (FA) was the first authority to change the rule of awarding two points for a win in the domestic league to three points. The objective of such change was to increase the risk-taking nature of the teams and, in turn, induce them to play for a win instead of a draw, making the games more competitive. It was observed that football was becoming less attractive, with teams playing for a draw instead of a win and, therefore, in most cases, being satisfied with points shared among them, which seemed like a cooperative outcome. The authorities suggested that changing the rule to three points for a win system would reduce the chance of a draw in the games and further increase the competitiveness among the teams.

In the paper, we intend to causally measure this rule change's impact on the domestic league's competitive balance using the quasi-experimental estimation methods. The contributions of our study are manifold. First, we employ a comparative study design and synthetic control method to analyze the impact of the three-point rule on competitiveness in the English domestic league. To our knowledge, only a few studies have employed causal estimation methods to this question. Further, no one has used the synthetic control method for this question, which is one of the best-fitted approaches for the question at hand. Second, instead of focusing on team-level outcomes, we intend to analyze the impact at the league level and, therefore, complement the existing studies focusing on team-level outcomes. We also provide a blueprint to potential researchers on how the synthetic control method, used mainly in natural experiment scenarios in economics and political science literature, can be applied to sports analytics.

Before discussing our findings, we summarize the existing studies and their findings. The impact of the three-point rule is well analyzed in the context of domestic football across Europe. Dilger and Geyer (2009) use the variation between league games versus cup competition in German football before and after the three-point rule change to analyze the impact on the number of draws and wins in the German football league. Using a difference-in-difference (DID) approach, they estimate the effect of the three-point rule and find positive evidence of a reduction in the number of draws and an increase in the number of wins. However, this study suffers from a significant drawback. The teams participating in the cup competition and German league overlap with each other, which leads to contamination of the control group, and would bias the impact estimated. Further, this design also violates the crucial assumption of parallel trends needed for implementing DID approach to the problem. Dewenter and Namini (2013) provide a strategic model for analyzing the impact of three-point rule and use the data from the German league to conclude that due to a change in the three-point rule, the home team's goal-scoring and number of wins decreased. In contrast, the reverse happens for the away teams. Hon and Parinduri (2014) find that the three-point rule induces the lagging teams to perform more aggressively in the second half, whereas there is no change in the strategy of the leading teams. In the Portuguese league, Guedes and Machado (2002) find that the three-point rule does not affect teams' attacking behaviour except for weak teams that play more defensively against strong opponents. Moschini (2010), using data from 35 countries and 30 years and applying panel data models, showed that effect of the three-point rule change has been significant in terms of increasing the number of goals and reducing the number of draws.

Within this literature, there has yet to be a consensus on the impact of change in the three-point rule on the match outcomes. One reason for this could be that the methodologies used by the researchers are not meant to capture the causal effect of the rule change, except in the case of Dilger and Geyer (2009). Second, none of these studies focus on finding a suitable control group but use time series variations in the match outcomes over the years/seasons to measure the effect. Lastly, none of these studies analyzes the impact of three-point rule on the macro indicators at the league level, such as competitive balance[1], dispersion in points scored in the league and so on. Notably, a few of the studies (Moschini, 2010) have computed some basic indicators, such as the number of draws and wins (without accounting for the total matches played) to measure league-level performance changes. We attempt to complement these studies by providing a more macro analysis and introducing a novel quasi-experimental approach to answer this question: What is the impact of three-point rule change on the football leagues and their competitiveness?

[1] See Goossens (2006) and Dobson and Goddard (2011) for a detailed discussion on competitive balance indicators in football. A detailed discussion on them is provided in online appendix A1.

For our analysis, we use data from six leagues (English, German, Spanish, Dutch, Italian and French) from 1963 to 1993. First, using the synthetic control method, we find that a three-point rule change leads to an increase in competitive balance in the league. Second, we show that these findings are robust to various placebo tests and leave-one-out test. Third, we show that similar results emerge by using alternate measures of competitive balance, the $\widehat{NAMSI}$ index. Fourth, we use the Difference-in-Differences (DID) method to estimate the impact of three-point rule change on the Distance to Competitive Balance (DCB) measure of competition in the league and find comparable results. Lastly, we also checked whether there was an increase in the number of goals scored per match after the three-point rule change and found no significant impact.

The rest of the paper is structured as follows. In the next section, we discuss the data, followed by details of various measures of competitiveness. In section 3, we outline the synthetic control method's design. Section 4 has the main results and robustness checks. Section 5 concludes.

## 2. Data and Measures

### 2.1 Data used

For our analysis, we use data from six major domestic leagues in Europe that is English (First Division/ Premier League), German (Bundesliga), Spanish (La Liga), Dutch (Eredivisie), Italian (Serie A) and French (Ligue 1), for the period 1963 (season 1963-64) to 1993 (season 1993-94). We have the match-level data for these seasons and leagues, which we use to create league-level indicators of competitiveness. We restrict our data to season 1994-95 because, before that season, only the English league had adopted the three-point rule, whereas none of the other leagues had moved to this rule. In England, the three-point rule was introduced in the season 1981-82. For other leagues, the years were German in 1995-96, Spanish in 1995-96, Dutch in 1995-96, Italian in 1994-95 and French in 1994-95.

Thus, the period from 1963 to 1993 provides a natural experiment setting, where only one group of observations (English League in our case) is given the treatment of the three-point rule. In contrast, no other potential control groups have been exposed to this treatment. The other five leagues can act as potential control group. Additionally, a long period before the treatment, 1963-1980, provides a controlled setting where treatment and control groups can be compared to find the baseline differences, if any, and then select the correct control group.

Before defining the indicators of competitive balance, a few caveats are in place. First, as we move from team/match to league-level analysis, the numbers of correlates become very limited, a well-accepted limitation of macro-level comparative analysis. Second, the cross-sectional variation in outcomes becomes limited; therefore, the panel data method should be used cautiously.

### 2.2 Measures of competitive balance

In the past few decades, relevant literature has extensively developed on measuring competitive balance in sports, especially football. In this section, we will restrict the discussion to measuring competitive balance without engaging in a detailed literature review for brevity. Some of the key competitive balance measures recommended in football are a) competitive inequality measure based on the deviation from the draw, b) relative competitive inequality accounting for the number of matches, c) national measure of seasonal imbalance (NAMSI), d) NAMSI adjusted for the share of average wins in the season (See Goossens, 2006; Dobson & Goddard, 2011). Most of these measures are widely used in football and other sports to measure the nature of competitive balance in the season. Another set of measures is adopted from industrial organization and market concentration literature. Hirschman-Herfindahl Index (HHI) measures the concentration of market power in the industrial organization literature. Variants of HHI have been proposed in sports literature,[2] such as a) HHI, where the share of wins for each team is calculated, b) Adjusted

[2] Please refer to Owen (2007) for a detailed discussion on HHI-based measures of competitive balance.

HHI for the lower limit measured by 1/K where K = number of teams, and c) a normalized HHI using lower and upper bounds. For a detailed discussion on both sets of measures, please refer to the online appendix A1 to the paper.

Though all these measures are helpful and try to capture competitive balance in football competition, we focus on a more refined measure in the family of HHI indices proposed by Triguero Ruiz and Avila-Cano (2019). This measure is termed "Distance to Competitive Balance" (DCB). This measure provides the ordinality across the levels of competitiveness and a cardinal comparison across the leagues and seasons. Further, it suffices with some of the axiomatic requirements of the inequality measurement indices[3]. The formula for the measurement of DCB is as follows:

$$\text{DCB (s)} = \sqrt{HHI_{norm}} = \sqrt{\frac{HHI - HHI_{min}}{HHI_{max} - HHI_{min}}}$$

Where $HHI = \sum_i^K {s_i}^2$ and $s_i$ is the share of each team's total points in a season/league. $HHI_{min}$ is the value of this index generated by the league configuration of minimum concentration, where all the teams draw their matches with all other teams, leading to a uniform distribution equal to 1/I, where I is the number of teams in the league. $HHI_{max}$ is the value of index with the maximum concentration of league.[4]. This process leads to creating an index with a natural interpretation. If the value of the index tends towards 1, that is the maximum concentration of the league, whereas 0 leads to the minimum concentration. Further, this normalization leads to ease of comparison across time and leagues and comparison of leagues with different formations in terms of the number of teams and matches played.

[3] For a detailed discussion on the properties of DCB, please refer to Triguero Ruiz and Avila-Cano (2019).
[4] For the upper and lower bounds of HHI in the football competition, please refer to Avila-Cano et al. (2021).

In our context, these properties of DCB become helpful for multiple purposes. First, in our context, we compare the outcomes over time and across various leagues, where DCB will be helpful due to the natural bounds of this index. Second, before and after the three-point rule, there will be changes in the point distribution across the teams, which can affect inequality. DCB measures remain unaffected by such changes and will be comparable between the leagues with a three-point and two-point rule. Lastly, interpreting results and changes in the DCB index is natural to explain.

In the next section, we discuss the estimation strategy for our analysis.

## 3. Estimation strategy and specification

The main aim of this paper is to assess the impact of the three-point rule change on the competitive balance in the English top league (earlier called Division One and now Premier League) to be measured in terms of DCB.

We need a control group to estimate the causal impact of the rule change on the English league. During the period before the rule change, the competitive nature of the English league could have been very different from that of other leagues. Therefore, direct comparison with other leagues (where the rule was not changed) may not be helpful.

We tackle this problem by comparing the temporal evolution of the English league with that of a weighted average of comparable leagues from other countries. Given that we require a long time series of league performances for the control group, we use data from the top five domestic leagues from European football (German, Italian, Dutch, Spanish and French). We will call this weighted average of other leagues a synthetic English league without the rule change, and this would act as a comparison (control) group for the actual English league with the rule change. The discussion here on, for the synthetic control method, is based mainly on Abadie and Gardeazabal (2003) and Abadie et al. (2010).

Let $I$ be the number of leagues available as a control group and $G = (g_1, g_2, \dots g_I)$ be an $(I \times 1)$ vector of non-negative weights with sum adding up to one. Here, $g_i$ represents the weight assigned to league $i$ in the synthetic English league. Each vector of G produces a different synthetic English league; therefore, the optimal choice of these weights is essential to create the correct control group. Such weights are chosen so that the synthetic English league resembles the English league before the three-point rule change.

Let $X_1$ be a $(K \times 1)$ vector of indicators before the three-point rule change for the English league. Also, $X_0$ be a $(K \times I)$ matrix which contains values of the same indicators for the potential control group football leagues (in our case, other countries' leagues). Let V be a diagonal matrix with non-negative components. The values of the diagonal elements of V reflect the relative importance of the different indicators. The vector of weights G* is chosen to minimize $(X_1 - X_0G)'V(X_1 - X_0G)$ subject to $G > 0$ $(i = 1, 2, \dots, I)$ and $\sum_i^I g_i = 1$. The vector G* defines the combination of non-rule change control leagues that best resemble the English league in terms of performance indicators at the outset of the three-point rule change for the English league.

We have a $Y_1(T \times 1)$ vector whose values are the values of various football league indicators over the time periods (seasons), T. We also have the corresponding values for the control group leagues, say $Y_0(T \times J)$ matrix. We aim to estimate the comparable values for the synthetic English league using the optimal weight generated from the earlier optimization, $Y_1^* = Y_0G^*$. If the effect of the three-point rule change is there, we should observe that values of $Y_1$, $Y_1^*$ would diverge after the treatment period, that is, 1980.

### *3.1 Choice of matching variables for synthetic control*

It is one of the most critical steps in creating synthetic control for the treatment units. Abadie (2021) provides some guidance in the variable selection process and suggests that pre-intervention values of the outcome variable in panel data are crucial for synthetic control, in line with the vector auto-regression model. Further, Abadie (2021) suggests that variables that co-move in time across countries can also be good contenders for the matching variables. We consider the average share of wins and draws in a season and the number of league teams to be indicators. Further, Ferman et al. (2020) suggest that many different specifications should be shown as robustness checks in case of no consensus on a single specification. Our analysis considers various specifications and shows their range of average treatment effects[5].

### *3.2 Robustness and sensitivity analysis of synthetic control*

Abadie *et al.* (2015) and Abadie (2021) suggest that there are two ways in which the robustness of the estimates can be tested. One is to question the study's design regarding the choice of the donor pool and the second is in terms of the choice of predictors of the outcome variables. These studies suggest that the leave-one-out robustness test helps analyze the sensitivity of the estimate and their robustness by leaving one of the donor pool choices out at a time. This will provide a benchmark regarding the treatment effect size in the wake of suboptimal synthetic control. Please refer to Abadie et al. (2015) for a detailed discussion. Another test on the choice of matching variables is already discussed in the previous subsection. Further, we also conduct various placebo studies to test the robustness of our model in line with Abadie et al. (2015) and Abadie (2021).

In the next section, we provide our main results.

[5] For brevity, the results of additional specifications are presented in the online appendix A2. We are grateful to one of the anonymous reviewers for recommending this part.

## 4. Estimation results

There are two steps in estimating the average treatment effect of the three-point rule on competitive balance for the English top-tier league. In the first step, we construct a synthetic English league for our counterfactual so that the synthetic English league reproduces the value of the English league in the three-point rule era, if rule would not have changed. Next, we estimate the rule change's Average Treatment Effect (ATE) on the competitive balance in English top-tier football.

### *4.1 Constructing a synthetic version of the English league*

For the construction of the synthetic version of the English league, we estimate the weights using the following predictors: average share of wins and draws, number of teams count, and all two-period lag values of DCB (that is, $t_0, t_2, t_4$ and so on). Next, based on the minimization of the root mean squared error (RMSE), we choose the specification, where the weights chosen for the predictors in descending order are as follows: DCB (1969) 0.3485, DCB (1965) 0.2119, average share of wins 0.1587, number of teams count 0.1544 and average share of draws 0.0777 and DCB (1979) 0.0488. A detailed description of the same is available in Table 1.

[Insert table 1 here]

Next, we check the weights assigned to the leagues in the donor pool to construct the synthetic English League. We observe positive weights of three leagues: French 0.6010, Spanish 0.3350, and Dutch 0.0650. Our choice of specification with these weights is based on minimization RMSE across the specification (Abadie, 2021; Ferman *et al.*, 2020). As Ferman *et al.* (2020) suggested, we also provide the details of alternate specifications and their findings in the online appendix A2. We consider three other specifications: five-period gap lags, three-period gap lags, and one-period gap all lags. In these specifications, we find that within the donor pool, positive weights are assigned to French, Italian, Dutch, and Spanish (one-period lags model), French and Dutch (three-period lags model), and French, Italian, and Dutch (five-period lags model).

Interestingly, French and Dutch leagues always remain in all specifications. It is tough to explain the reason for the choice of French and Dutch leagues in creating a synthetic English league, as they are considered lower level and different from the English league. Based on the league-level statistics, one can argue that in terms of competitiveness among the teams in the league, they were very similar to the English league during the pre-intervention period.

### *4.2 The effect of three-point rule on competitive balance*

Having obtained the synthetic English league through a linear combination of other leagues, we now compare the outcome in the post-intervention period to measure the average treatment effect of the three-point rule on competition balance in the English League.

Before discussing the results, it is essential to understand how to interpret the findings from the DCB measure of competitive balance. Triguero Ruiz and Avila-Cano (2019) explain that a higher value of DCB refers to an increased imbalance in the league and a lower value indicates an increase in competitive balance in the league. Further, given that the range of the measure is between 0 and 1, we can interpret it in terms of points reduction. For example, suppose the value of DCB for a league changes from 0.40 to 0.30. In that case, compared to the maximum concentration, there is a 10-point increase in competitive balance or a reduction in league concentration. Having said so, we move on to our findings.

[Insert table 2 here]

As highlighted in Table 2, we find that in the period 1981-1993, there was an increase in the competitive balance in English league, as measured by the average treatment effect of -0.051. This indicates that due to the introduction of the three-point rule, there is an increase in competitive balance in the English league. One plausible explanation for this is as follows. Due to the increase in the points for a win, the teams are likely to be more aggressive, which explicitly benefits lower-end teams because of the tilted rewards for aggressive play. This increases the possibility of better dispersion of points and results, leading to more exciting leagues and, in turn, improving the competitive balance. Further, this was one of the critical goals of the rule change in the first place. In Figures 1 and 2, we show the same results graphically. Figure 1 shows the trajectory of DCB for both the treatment unit (English league) and control unit (synthetic English league) before and after the intervention of the three-point rule. In Figure 2, we show the treatment effect, which shows a systematic reduction in competitive concentration in the league (increase in competitive balance).

[Insert figure 1 here]

[Insert figure 2 here]

Next, we briefly discuss the findings of alternate specifications. We consider three other specifications with the five-period gap lags, three-period gap lags, and one-period gap. We find that the average treatment effect in these three specifications remains in the range of -0.0571 to -0.073. This indicates very similar findings across the specification and precisely the direction of impact towards an increase in competitive balance and reduction in concentration. For more details on the findings from alternative specifications, please refer to online appendix A2.

### *4.3 Placebo study*

In this sub-section, we check the robustness of our findings in terms of increased competitive balance using a placebo test. We assign the treatment to one of the leagues from the donor pool and then estimate the average treatment effect (Abadie et al., 2015). Ideally, the effect on the non-treatment country should be close to nil (insignificant). Further, this effect helps us calculate the potential bias in our estimates due to some unobserved heterogeneity. Assigning the placebo treatment to Dutch, Spanish, French, German, and Italian leagues, the average treatment effect (ATE) is 0.0097, 0.0124, 0.0026, -0.0219, and 0.0145, respectively. This highlights that the effect on the placebo leagues is negligible and the estimates for the English league are robust and not due to some randomness.

Further, we also perform a placebo test by changing the year of the three-point rule change from 1981 to 1969 and see the impact of this placebo treatment on the outcome from 1969 to 1981. ATE is very minuscule at -0.0175, indicating that the main effect is robust and significant compared to any random effect.

Additionally, as suggested by Abadie (2021) and Abadie *et al.* (2015), we also conduct a robustness test termed as leave-one-out test to check the sensitivity of our estimates to change in the weight assigned to various leagues from our donor pool. In this test, we randomly leave one of the leagues from the chosen leagues (that is, French, Spanish, and Dutch) and estimate the synthetic English league again to measure the effect of the rule change. The test results are seen in Figure 3, showing that our results remain unchanged in terms of the effect on DCB due to rule change. In Table 3, we show the average treatment effect in case of leaving-one-out of the test and find that they are in line with our main findings.

[Insert figure 3 here]

[Insert table 3 here]

### 4.4 *Estimates using the difference-in-differences (DID) method*

An alternative methodology to be employed in this context, which is similar to the Synthetic Control Method (SCM) is Difference-In-Differences (DID)[6]. The critical distinction is that in SCM, we use the weight generated through the minimization of root mean squared errors; whereas, in DID, we use equal weights for all the donor pool control groups and then estimate the ATE. In our context, when we estimate ATE through DID, we find that due to a three-point rule change, there is a decrease in concentration and an increase in competitive balance. The ATE estimate for the DID model is -0.0545 (see Table 4), which is very similar to our estimates of SCM.

[Insert table 4 here]

### *4.5 Additional results using alternate measures of league performance*

In this sub-section, we provide ATE for alternate measures of league performance. First, we use the average number of goals scored per match. The main reason for doing this analysis is as follows. The three-point rule would affect the risk-taking chances of both home and away teams to play more aggressively; however, this may lead to a decrease in chances of draw. Given this discussion, in either of the instances, there is a possibility that one outcome is in the control of the team and likely to increase in the number of goals scored in a match. For the sake of brevity, we do not report the full estimates and results in the paper[7]. We find that there is no significant change in the number of goals scored in the English league due to the three-point rule. The ATE is -0.0131.

---

[6] For the sake of brevity, we do not indulge in a detailed discussion on the DID method. For a more detailed discussion on the same in the context of football, please refer to Dilger and Geyer (2009).
[7] The full estimates are available in the online appendix A3.

Next, we use $\widehat{NAMSI}$ as a measure of seasonal imbalance.[8] Using this indicator, we also find a decrease in the imbalance in the league, or in other words, there is an increase in competitive balance in the English league after the three-point rule. The ATE for the model is -0.0297.

In summary, there is an increase in the competitive balance in the English league due to the introduction of the three-point rule.

## 5. Conclusion

This paper provides an alternative methodology of the synthetic control method, adapted from causal inference literature, to measure the impact of three-point rule change on competitive balance in the league. To the best of our knowledge, this is the first study to utilize SCM methodology in the context of football. The main findings of this paper can be summarized as follows. First, the three-point rule change increases competitive balance, measured through DCB for the English League. Second, we show that these findings are robust to various placebo tests and leave-one-out test. Third, we show that similar results emerge by using alternate measure of competitive balance, the $\widehat{NAMSI}$ index. Fourth, we use the DID method to estimate the impact of three-point rule change on DCB and find comparable results. Lastly, we also checked whether there was an increase in the number of goals scored per match after the three-point rule change and found no significant results.

This study helps academics and sports authorities understand the various methodologies to be adopted for evidence-based policymaking and can further enrich the nature of competition and its determinants in the short and long run. This study increases the toolbox of sports enthusiasts, academics, and regulators by providing a comparative study measure. Further, we also provide the causal analysis of the sports policy on the league-level outcomes, a novel contribution to this literature.

---

[8] For a discussion on the measurement of NAMSI2, please refer to online appendix A1; for estimation results, refer to online appendix A4.

It is essential to say that no study is devoid of any shortcomings. One of the primary shortcomings of SCM is that it is limited primarily to applications at the macro/aggregate level. This means it will not be suitable for a micro-level analysis at the team level. In future studies, researchers can attempt to combine SCM with DID to understand the impact of such rule change at the team (micro) level.

# Tables

Table 1: Covariate balance in the pretreatment period

| Covariate | V.weight | Treated | Synthetic value | Control bias | Average value across donor pool | Control bias |
|---|---|---|---|---|---|---|
| DCB(1963) | 0 | 0.323 | 0.2923 | -9.49% | 0.3676 | 13.81% |
| DCB(1965) | 0.2119 | 0.3586 | 0.406 | 13.22% | 0.4406 | 22.87% |
| DCB(1967) | 0 | 0.3346 | 0.3321 | -0.75% | 0.3709 | 10.86% |
| DCB(1969) | 0.3485 | 0.4129 | 0.3922 | -5.03% | 0.4203 | 1.78% |
| DCB(1971) | 0 | 0.409 | 0.371 | -9.29% | 0.4516 | 10.41% |
| DCB(1973) | 0 | 0.2842 | 0.3108 | 9.37% | 0.3684 | 29.66% |
| DCB(1975) | 0 | 0.3718 | 0.3047 | -18.04% | 0.3547 | -4.60% |
| DCB(1977) | 0 | 0.4095 | 0.3406 | -16.83% | 0.3664 | -10.53% |
| DCB(1979) | 0.0488 | 0.3367 | 0.4173 | 23.96% | 0.3785 | 12.41% |
| Average share of wins in a season | 0.1587 | 0.3637 | 0.3672 | 0.96% | 0.3576 | -1.67% |
| Average share of draws in a season | 0.0777 | 0.2726 | 0.2664 | -2.27% | 0.2847 | 4.44% |
| Number of teams in a season | 0.1544 | 21.774 | 19.108 | -12.24% | 18.0387 | -17.16% |

Source: Author's calculations using the data

"V.weight" is the optimal covariate weight in the diagonal of the V matrix.
"Synthetic Control" is the weighted average of control units in the donor pool with optimal weights.
"Average Control" is the simple average of control units in the donor pool with equal weights.

Table 2: Average treatment effect (ATE) and prediction results in the three-point rule period

| Time | Actual Outcome | Predicted Outcome | Treatment Effect |
|---|---|---|---|
| 1981 | 0.35 | 0.44 | -0.10 |
| 1982 | 0.25 | 0.36 | -0.11 |
| 1983 | 0.30 | 0.40 | -0.10 |
| 1984 | 0.36 | 0.37 | -0.01 |
| 1985 | 0.43 | 0.40 | 0.04 |
| 1986 | 0.31 | 0.41 | -0.10 |
| 1987 | 0.41 | 0.30 | 0.11 |
| 1988 | 0.32 | 0.49 | -0.16 |
| 1989 | 0.32 | 0.35 | -0.03 |
| 1990 | 0.38 | 0.37 | 0.02 |
| 1991 | 0.30 | 0.36 | -0.07 |
| 1992 | 0.25 | 0.40 | -0.15 |
| 1993 | 0.38 | 0.39 | -0.01 |
| **Mean (ATE)** | **0.34** | **0.39** | **-0.05** |

Source: Author's calculations using the data

Table 3: Leave one out Test: Minimum and Maximum Synthetic outcomes

| Time | Treatment Effect | Treatment Effect (LOO) | |
|---|---|---|---|
| | | Min | Max |
| 1981 | -0.10 | -0.12 | -0.06 |
| 1982 | -0.11 | -0.18 | -0.08 |
| 1983 | -0.10 | -0.15 | -0.08 |
| 1984 | -0.01 | -0.04 | 0.01 |
| 1985 | 0.04 | -0.04 | 0.08 |
| 1986 | -0.10 | -0.12 | -0.05 |
| 1987 | 0.11 | 0.03 | 0.14 |
| 1988 | -0.16 | -0.18 | -0.11 |
| 1989 | -0.03 | -0.12 | 0.01 |
| 1990 | 0.02 | -0.02 | 0.04 |
| 1991 | -0.07 | -0.11 | -0.07 |
| 1992 | -0.15 | -0.19 | -0.12 |
| 1993 | -0.01 | -0.03 | 0.01 |

Source: Author's calculations using the data

| Table 4: Difference-in-differences estimates for Distance to Competitive Balance (DCB) | |
|---|---|
| Explanatory variables | Outcome variable: DCB |
| Time Dummy | 0.0237* |
| | (0.0132) |
| Time*Treatment | -0.0545** |
| | (0.0241) |
| Treatment | 0.0168 |
| | (0.0234) |
| Average share of wins in a season | -24.28*** |
| | (4.125) |
| Average share of draws in a season | -12.17*** |
| | (2.078) |
| Number of teams in a season | -0.0109*** |
| | (0.00390) |
| Constant | 12.73*** |
| | (2.104) |
| Observations | 186 |
| R-squared | 0.101 |

Robust standard errors in parentheses
*** p<0.01, ** p<0.05, * p<0.1
Source: Author's calculations using the data

Figures

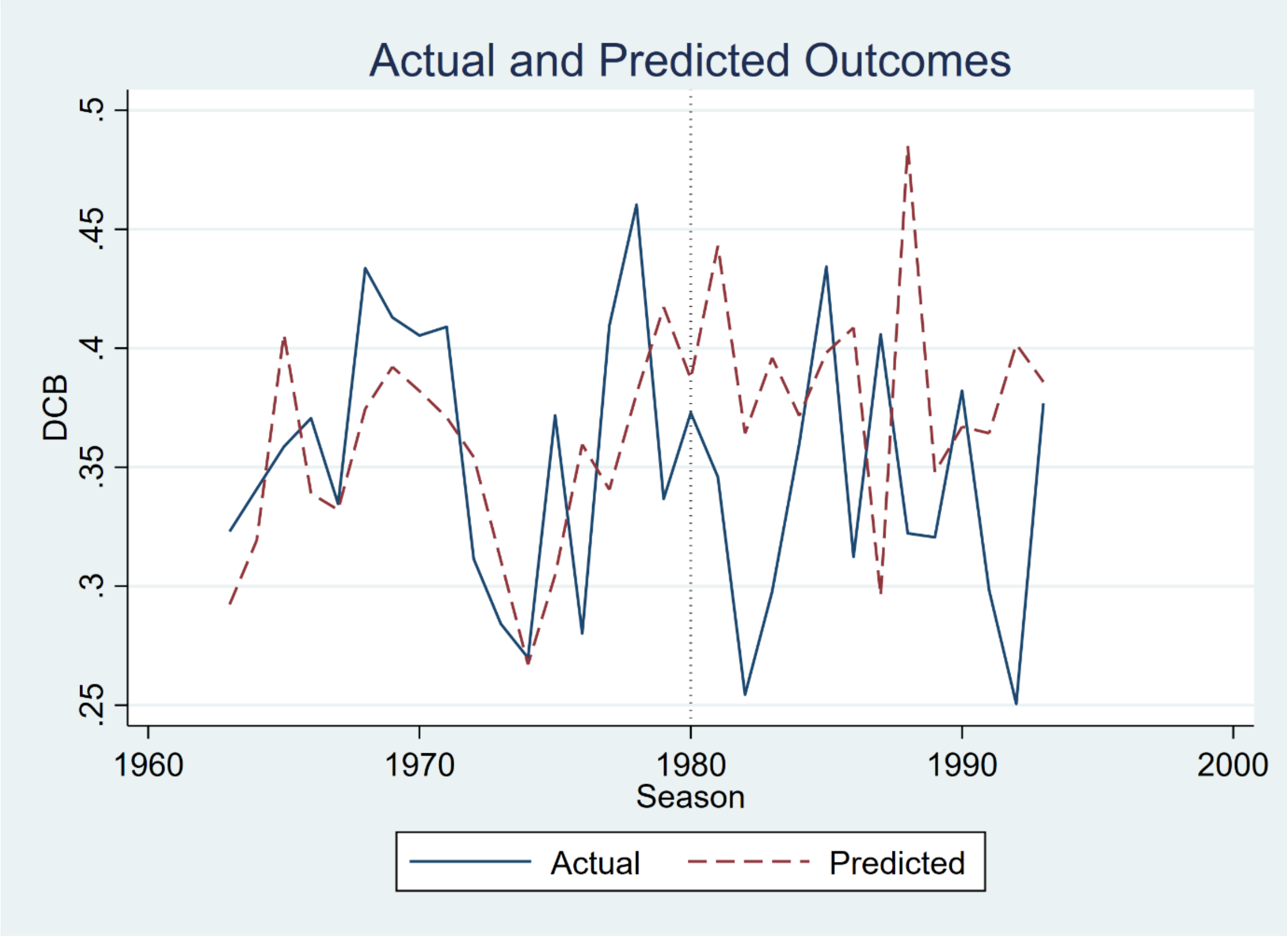

Figure 1: Actual versus predicted (Synthetic) Outcome (Outcome Variable: Distance to Competitive Balance (DCB))

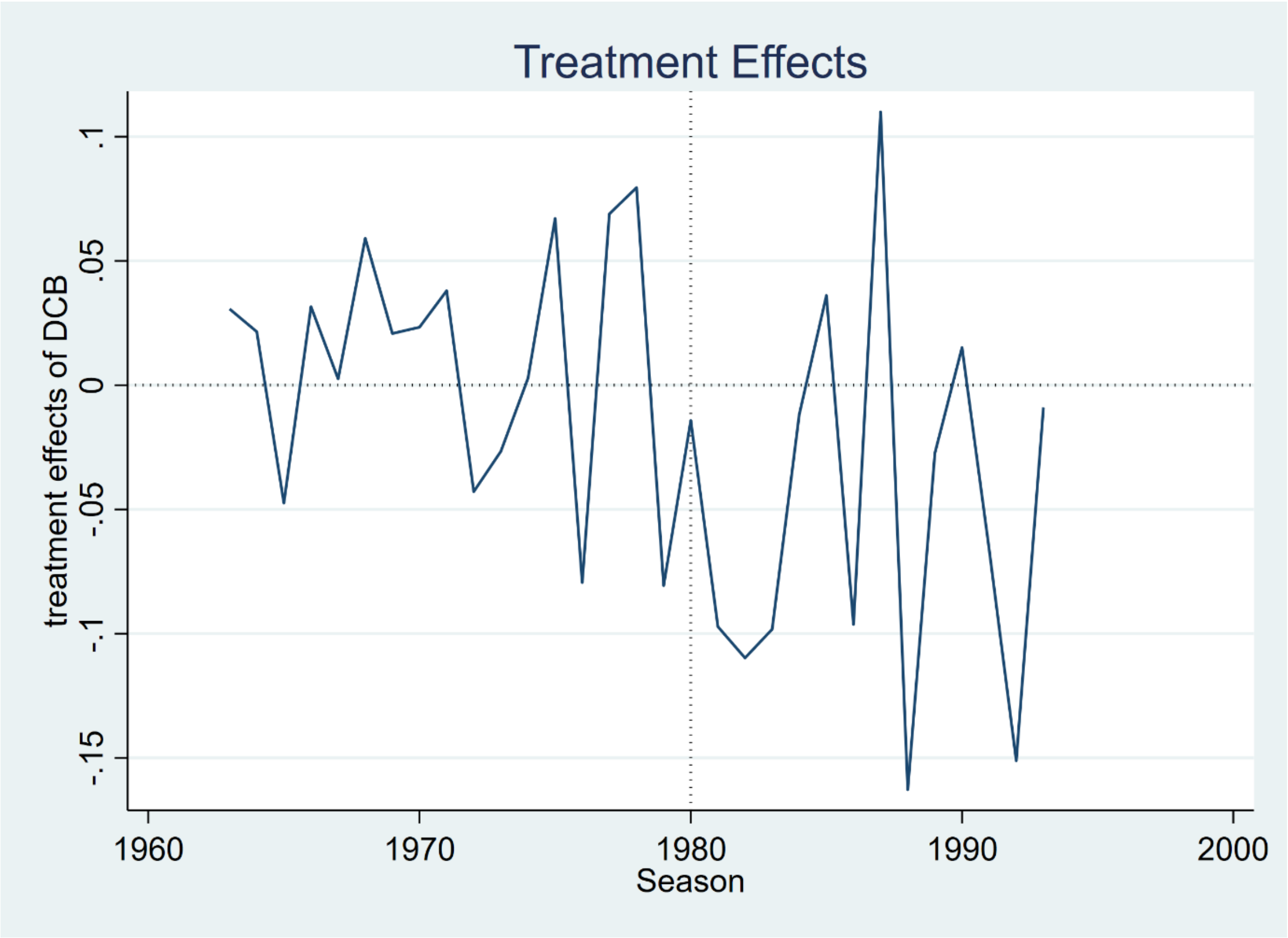

Figure 2: Treatment effect of three-point rule (Outcome variable: Distance to Competitive Balance (DCB))

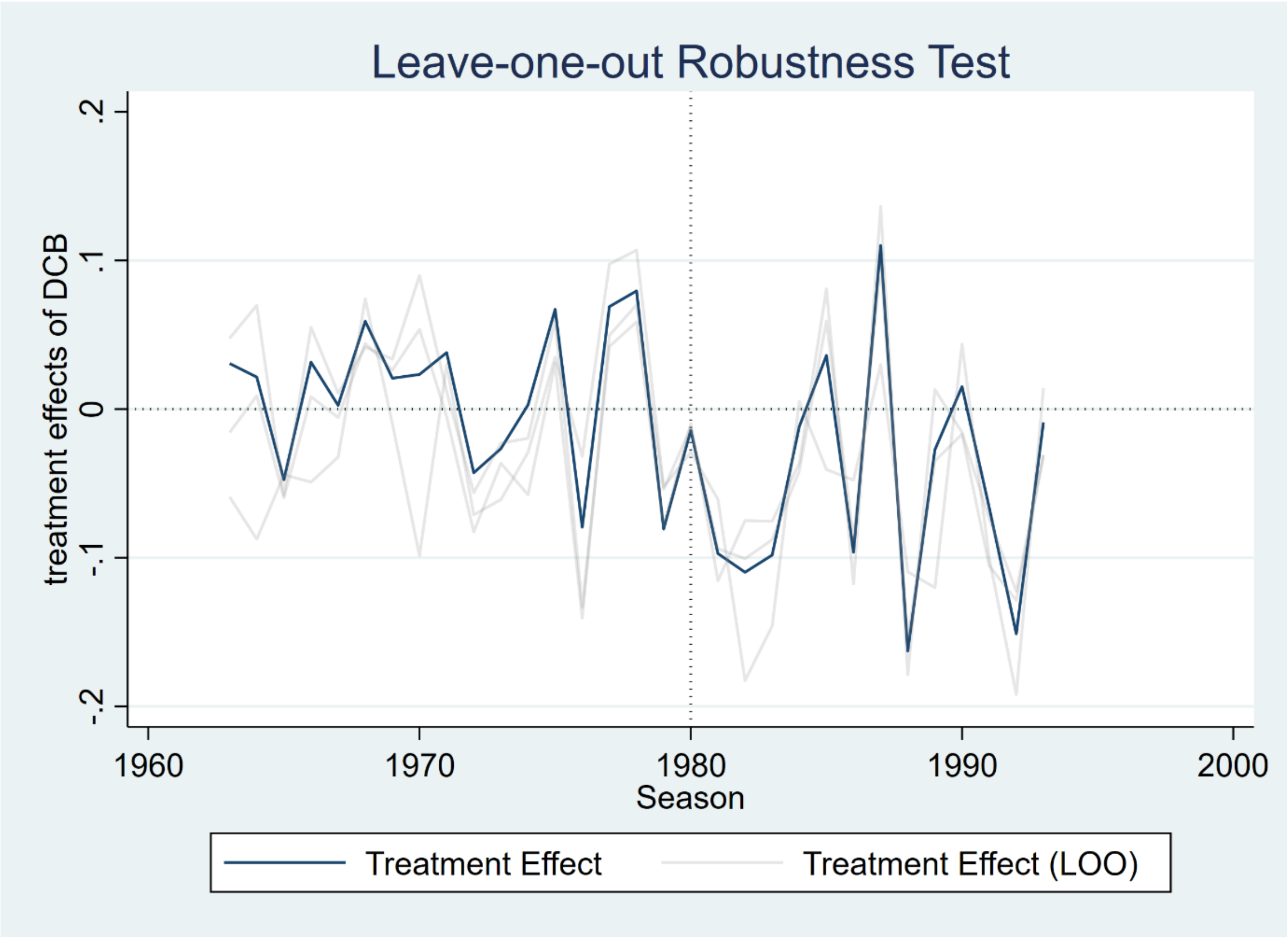

Figure 3: Leave-one-out Robustness Test (Outcome: Average number of goals scored per team)

Appendix A

***Appendix A1: Discussion on various measures of competitive balance***

The first measure of competitive inequality is the win percentage dispersion among the league teams for a season. Suppose there are no draws/ties; all matches lead to a winner and a loser. In that case, the primary indicator of competitive inequality can be computed as $\sigma_{it} = \sqrt{\frac{\sum_1^k (w_k - 0.5)^2}{k}}$, where i= country, t = season and k= teams in the league of country $i$. Based on this indicator, a higher value of $\sigma_{it}$ reflects a higher level of inequality within the league. One drawback of this indicator is that it depends on the number of teams and matches played, which do not remain constant within a league over time. Therefore, a relative competitive inequality measure is proposed in the literature, that is, $r_{it} = \sigma_{it} / (0.5/\sqrt{m})$, where m is the number of matches played.

Having discussed these two measures, we propose another variant for the analysis of football matches where not all matches lead to a win or loss but may end up in a draw. Instead of using value half for the mean numbers of wins in a season under perfect equality, in modified variations of the above-proposed measures, we use the statistics of the mean realized number of wins in a season for our analysis. This leads to two additional indicators. Further, the difference between these variants also helps us analyze the impact of the share of draws on the competitive inequality measures. We call the first variant as $\widehat{\sigma_{it}} = \sqrt{\frac{\sum_1^k (w_k - \overline{w_{it}})^2}{k}}$, where $\overline{w_{it}}$ is the mean share of wins in season t for league i. The second variant is $\widehat{r_{it}} = \widehat{\sigma_{it}} / (0.5/\sqrt{m})$.

The next indicator of competitive (im-)balance is proposed by Goossens (2006), which focuses on complete predictability as a benchmark while calculating inequality. The national measure of seasonal imbalance (NAMSI) is as follows: $NAMSI = \sqrt{\frac{\sum_1^k (w_k - 0.5)^2}{\sum_1^k (w_{i,max} - 0.5)^2}}$. This measure takes the value one if there is complete predictability in the games, i.e., the top team wins all the matches, whereas the second team wins all the matches except the one with the top team and so on; it takes the value zero if each team achieves the wins share of half each. Along the lines of the variants proposed for the earlier measures, we also have a variant for NAMSI: $\widehat{NAMSI} = \sqrt{\frac{\sum_1^k (w_k - \overline{w_{it}})^2}{\sum_1^k (w_{i,max} - \overline{w_{it}})^2}}$.

Lastly, we also calculate the Hirschman-Herfindhal index (HHI) for the inequality in the share of wins and draws by the teams in a season. In an unequal league, there should be a concentration of wins, whereas in a league with equality, there should be more draws. Additionally, we measure a variant of HHI that accounts for the number of teams in the league while calculating HHI. The formula for these two measures is as follows. $HHI_W = \sum_1^k (w_k)^2$ , where $w_k$ is the share of wins for each team in a season. The adjusted indicator would be $AHHI_W = \sum_1^K (w_k)^2 - 1/K$ , where $w_k$ is the share of wins for each team in a season. Similarly, we can calculate the HHI for the share of draws - $HHI_D = \sum_1^k (d_k)^2$ and $AHHI_D = \sum_1^K (d_k)^2 - 1/K$

## Appendix A2: Alternative specifications for synthetic English league and estimation using SCM

### *Specification 1: Using all the one-period gap lags of DCB for the construction of a synthetic English league.*

*Constructing a synthetic version of the English league*

For the construction of the synthetic version of the English league, we estimate the weights using the following predictors: average share of wins and draws, number of teams count, and all one-period lag values of DCB (that is, $t_0, t_1, t_2$ and so on). Table A1 compares treated and synthetic control values for all matching covariates.

| Table A1: Covariate balance in the pretreatment period | | | | | | |
|---|---|---|---|---|---|---|
| Covariate | V.weight | Treated | Synthetic value | Control bias | The average value across donor pool | Control bias |
| DCB(1963) | 0 | 0.323 | 0.34 | 5.27% | 0.3676 | 13.81% |
| DCB(1964) | 0 | 0.3408 | 0.3322 | -2.54% | 0.344 | 0.92% |
| DCB(1965) | 0.2209 | 0.3586 | 0.423 | 17.95% | 0.4406 | 22.87% |
| DCB(1966) | 0 | 0.3706 | 0.3871 | 4.45% | 0.3802 | 2.59% |
| DCB(1967) | 0 | 0.3346 | 0.3616 | 8.06% | 0.3709 | 10.86% |
| DCB(1968) | 0 | 0.4336 | 0.409 | -5.68% | 0.3717 | -14.29% |
| DCB(1969) | 0 | 0.4129 | 0.4163 | 0.81% | 0.4203 | 1.78% |
| DCB(1970) | 0.3636 | 0.4054 | 0.4153 | 2.45% | 0.4258 | 5.05% |
| DCB(1971) | 0 | 0.409 | 0.4336 | 6.01% | 0.4516 | 10.41% |
| DCB(1972) | 0 | 0.3114 | 0.4156 | 33.45% | 0.4203 | 34.96% |
| DCB(1973) | 0 | 0.2842 | 0.3556 | 25.13% | 0.3684 | 29.66% |
| DCB(1974) | 0 | 0.2699 | 0.3298 | 22.19% | 0.3716 | 37.68% |
| DCB(1975) | 0 | 0.3718 | 0.3557 | -4.33% | 0.3547 | -4.60% |
| DCB(1976) | 0 | 0.2802 | 0.3984 | 42.21% | 0.3823 | 36.46% |
| DCB(1977) | 0 | 0.4095 | 0.3655 | -10.75% | 0.3664 | -10.53% |
| DCB(1978) | 0.1951 | 0.4603 | 0.4014 | -12.80% | 0.3935 | -14.51% |
| DCB(1979) | 0.0652 | 0.3367 | 0.3946 | 17.20% | 0.3785 | 12.41% |
| Average share of wins in a season | 0.061 | 0.3637 | 0.3554 | -2.28% | 0.3576 | -1.67% |
| Average share of draws in a season | 0.0942 | 0.2726 | 0.288 | 5.65% | 0.2847 | 4.44% |
| Number of teams in a season | 0 | 21.7742 | 18.412 | -15.44% | 18.0387 | -17.16% |

Source: Author's calculations using the data
"V.weight" is the optimal covariate weight in the diagonal of the V matrix.
"Synthetic Control" is the weighted average of control units in the donor pool with optimal weights.
"Average Control" is the simple average of control units in the donor pool with equal weights.

Next, we check the weights assigned to the leagues in the donor pool to construct the synthetic English league. We observe that three leagues get positive weights: French 0.4070, Spanish 0.1910, Italian 0.2080 and Dutch 0.1930.

*The effect of the three-point rule on competitive balance*

Having obtained the synthetic English league through a linear combination of other leagues, we now compare the outcome in the post-intervention period to measure the average treatment effect of the three-point rule on competition balance in the English league.

As highlighted in Table A2, we find that in the period 1981-1993, there was an increase in the competitive balance in the English league, as measured by the average treatment effect of -0.0689.

Table A2: Average treatment effect (ATE) and prediction results in post three-point rule period

| Time | Actual Outcome | Predicted Outcome | Treatment Effect |
|---|---|---|---|
| 1981 | 0.35 | 0.45 | -0.11 |
| 1982 | 0.25 | 0.38 | -0.13 |
| 1983 | 0.30 | 0.42 | -0.12 |
| 1984 | 0.36 | 0.40 | -0.04 |
| 1985 | 0.43 | 0.41 | 0.02 |
| 1986 | 0.31 | 0.41 | -0.10 |
| 1987 | 0.41 | 0.33 | 0.07 |
| 1988 | 0.32 | 0.46 | -0.14 |
| 1989 | 0.32 | 0.37 | -0.05 |
| 1990 | 0.38 | 0.39 | -0.01 |
| 1991 | 0.30 | 0.42 | -0.12 |
| 1992 | 0.25 | 0.41 | -0.16 |
| 1993 | 0.38 | 0.40 | -0.03 |
| **Mean** | **0.34** | **0.40** | **-0.07** |

Source: Author's calligulations using the data

***Specification 2: Using all the three-period gap lags of DCB for the construction of a synthetic English league.***

*Constructing a synthetic version of the English league*

For the construction of the synthetic version of the English league, we estimate the weights using the following predictors: average share of wins and draws, number of teams count, and all three-period lag values of DCB (that is, $t_0, t_3, t_6$ and so on). Table A3 compares treated and synthetic control values for all matching covariates.

Next, we check the weights assigned to the leagues in the donor pool to construct the synthetic English league. We observe that three leagues get positive weights: French 0.6740 and Dutch 0.3260.

| Table A3: Covariate balance in the pretreatment period | | | | | | |
|---|---|---|---|---|---|---|
| Covariate | V.weight | Treated | Synthetic value | Control bias | The average value across donor pool | Control bias |
| DCB(1963) | 0.1193 | 0.323 | 0.289 | -10.53% | 0.3676 | 13.81% |
| DCB(1966) | 0.2161 | 0.3706 | 0.3586 | -3.24% | 0.3802 | 2.59% |
| DCB(1969) | 0 | 0.4129 | 0.4315 | 4.48% | 0.4203 | 1.78% |
| DCB(1972) | 0 | 0.3114 | 0.4254 | 36.61% | 0.4203 | 34.96% |
| DCB(1975) | 0 | 0.3718 | 0.3525 | -5.17% | 0.3547 | -4.60% |
| DCB(1978) | 0.3233 | 0.4603 | 0.4332 | -5.89% | 0.3935 | -14.51% |
| Average share of wins in a season | 0.0494 | 0.3637 | 0.364 | 0.07% | 0.3576 | -1.67% |
| Average share of draws in a season | 0.0596 | 0.2726 | 0.2718 | -0.27% | 0.2847 | 4.44% |
| Number of teams in a season | 0.2323 | 21.774 | 19.198 | -11.83% | 18.039 | -17.16% |

Source: Author's calculations using the data
"V.weight" is the optimal covariate weight in the diagonal of the V matrix.
"Synthetic Control" is the weighted average of control units in the donor pool with optimal weights.
"Average Control" is the simple average of control units in the donor pool with equal weights.

*The effect of the three-point rule on competitive balance*

Having obtained the synthetic English league through a linear combination of other leagues, we now compare the outcome in the post-intervention period to measure the average treatment effect of the three-point rule on competition balance in the English league.

As highlighted in Table A4, we find that in the period 1981-1993, there was an increase in the competitive balance in the English league, as measured by the average treatment effect of -0.0730.

| Table A4: Average treatment effect (ATE) and prediction results in post three-point rule period | | | |
|---|---|---|---|
| Time | Actual Outcome | Predicted Outcome | Treatment Effect |
| 1981 | 0.3459 | 0.4935 | -0.1476 |
| 1982 | 0.2544 | 0.3736 | -0.1192 |
| 1983 | 0.2978 | 0.4276 | -0.1298 |
| 1984 | 0.3598 | 0.4083 | -0.0484 |
| 1985 | 0.4343 | 0.4016 | 0.0327 |
| 1986 | 0.3124 | 0.4533 | -0.1409 |
| 1987 | 0.4058 | 0.3198 | 0.086 |
| 1988 | 0.3222 | 0.4713 | -0.1491 |
| 1989 | 0.3206 | 0.3218 | -0.0013 |
| 1990 | 0.3821 | 0.4027 | -0.0206 |
| 1991 | 0.2983 | 0.4096 | -0.1113 |
| 1992 | 0.2506 | 0.4127 | -0.1621 |
| 1993 | 0.3769 | 0.4143 | -0.0374 |

| **Mean** | **0.336** | **0.409** | **-0.073** |
|---|---|---|---|
| Source: Author's calculations using the data | | | |

***Specification 3: Using all the five-period gap lags of DCB for the construction of a synthetic English league.***

*Constructing a synthetic version of the English league*

For the construction of the synthetic version of the English league, we estimate the weights using the following predictors: average share of wins and draws, number of teams count, and all five-period lag values of DCB (that is, $t_0, t_5, t_{10}$ and so on). Table A5 compares treated and synthetic control values for all matching covariates.

Next, we check the weights assigned to the leagues in the donor pool to construct the synthetic English league. We observe that three leagues get positive weights: French 0.7020, Italian 0.1920 and Dutch 0.1050.

Table A5: Covariate balance in the pretreatment period

| Covariate | V.weight | Treated | Synthetic value | Control bias | The average value across donor pool | Control bias |
|---|---|---|---|---|---|---|
| DCB(1963) | 0 | 0.323 | 0.34 | 5.27% | 0.3676 | 13.81% |
| DCB(1964) | 0 | 0.3408 | 0.3322 | -2.54% | 0.344 | 0.92% |
| DCB(1965) | 0.2209 | 0.3586 | 0.423 | 17.95% | 0.4406 | 22.87% |
| DCB(1966) | 0 | 0.3706 | 0.3871 | 4.45% | 0.3802 | 2.59% |
| DCB(1967) | 0 | 0.3346 | 0.3616 | 8.06% | 0.3709 | 10.86% |
| DCB(1968) | 0 | 0.4336 | 0.409 | -5.68% | 0.3717 | -14.29% |
| DCB(1969) | 0 | 0.4129 | 0.4163 | 0.81% | 0.4203 | 1.78% |
| DCB(1970) | 0.3636 | 0.4054 | 0.4153 | 2.45% | 0.4258 | 5.05% |
| DCB(1971) | 0 | 0.409 | 0.4336 | 6.01% | 0.4516 | 10.41% |
| DCB(1972) | 0 | 0.3114 | 0.4156 | 33.45% | 0.4203 | 34.96% |
| DCB(1973) | 0 | 0.2842 | 0.3556 | 25.13% | 0.3684 | 29.66% |
| DCB(1974) | 0 | 0.2699 | 0.3298 | 22.19% | 0.3716 | 37.68% |
| DCB(1975) | 0 | 0.3718 | 0.3557 | -4.33% | 0.3547 | -4.60% |
| DCB(1976) | 0 | 0.2802 | 0.3984 | 42.21% | 0.3823 | 36.46% |
| DCB(1977) | 0 | 0.4095 | 0.3655 | -10.75% | 0.3664 | -10.53% |
| DCB(1978) | 0.1951 | 0.4603 | 0.4014 | -12.80% | 0.3935 | -14.51% |
| DCB(1979) | 0.0652 | 0.3367 | 0.3946 | 17.20% | 0.3785 | 12.41% |
| Average share of wins in a season | 0.061 | 0.3637 | 0.3554 | -2.28% | 0.3576 | -1.67% |
| Average share of draws in a season | 0.0942 | 0.2726 | 0.288 | 5.65% | 0.2847 | 4.44% |
| Number of teams in a season | 0 | 21.7742 | 18.412 | -15.44% | 18.0387 | -17.16% |

Source: Author's calculations using the data
"V.weight" is the optimal covariate weight in the diagonal of the V matrix.
"Synthetic Control" is the weighted average of control units in the donor pool with optimal weights.
"Average Control" is the simple average of control units in the donor pool with equal weights.

*The effect of the three-point rule on competitive balance*

Having obtained the synthetic English league through a linear combination of other leagues, we now compare the outcome in the post-intervention period to measure the average treatment effect of the three-point rule on competition balance in the English league.

As highlighted in Table A6, we find that in the period 1981-1993, there was an increase in the competitive balance in the English league, as measured by the average treatment effect of -0.0571.

Table A6: Average treatment effect (ATE) and prediction results in post three-point rule period

| Time | Actual Outcome | Predicted Outcome | Treatment Effect |
|---|---|---|---|
| 1981 | 0.35 | 0.45 | -0.11 |
| 1982 | 0.25 | 0.38 | -0.13 |
| 1983 | 0.30 | 0.42 | -0.12 |
| 1984 | 0.36 | 0.40 | -0.04 |
| 1985 | 0.43 | 0.41 | 0.02 |
| 1986 | 0.31 | 0.41 | -0.10 |
| 1987 | 0.41 | 0.33 | 0.07 |
| 1988 | 0.32 | 0.46 | -0.14 |
| 1989 | 0.32 | 0.37 | -0.05 |
| 1990 | 0.38 | 0.39 | -0.01 |
| 1991 | 0.30 | 0.42 | -0.12 |
| 1992 | 0.25 | 0.41 | -0.16 |
| 1993 | 0.38 | 0.40 | -0.03 |
| **Mean** | **0.34** | **0.40** | **-0.07** |

Source: Author's calculations using the data

## Appendix A3: Impact of the three-point rule on Average goals scored per match in the English league

*Constructing a synthetic version of the English league*

For the construction of the synthetic version of the English league, we estimate the weights using the following predictors: average share of wins and draws, number of teams count, and all two-period lag values of DCB (that is, $t_0, t_2, t_4$ and so on). Table A7 compares treated and synthetic control values for all matching covariates.

Next, we check the weights assigned to the leagues in the donor pool to construct the synthetic English league. We observe that three leagues have positive weights: French 0.3080, Spanish 0.2080 and Dutch 0.4830.

Table A7: Covariate balance in the pretreatment period

| Covariate | V.weight | Treated | Synthetic value | Control bias | The average value across donor pool | Control bias |
|---|---|---|---|---|---|---|
| Average goals scored per match(1963) | 0 | 1.2239 | 1.1672 | -4.64% | 1.0888 | -11.04% |
| Average goals scored per match(1965) | 0.2532 | 1.1486 | 1.1139 | -3.01% | 1.0404 | -9.42% |
| Average goals scored per match(1967) | 0.1626 | 1.1072 | 1.0174 | -8.12% | 0.9895 | -10.63% |
| Average goals scored per match(1969) | 0 | 0.9645 | 1.0118 | 4.91% | 0.9593 | -0.53% |
| Average goals scored per match(1971) | 0.4565 | 0.9206 | 0.9596 | 4.24% | 0.9415 | 2.27% |
| Average goals scored per match(1973) | 0 | 0.8738 | 1.0412 | 19.15% | 1.006 | 15.13% |
| Average goals scored per match(1975) | 0 | 0.9792 | 1.0328 | 5.47% | 1.0135 | 3.50% |
| Average goals scored per match(1977) | 0 | 0.98 | 1.0683 | 9.01% | 1.0282 | 4.92% |
| Average goals scored per match(1979) | 0 | 0.9197 | 1.0234 | 11.27% | 0.972 | 5.68% |
| Average share of wins in a season | 0.05 | 0.3637 | 0.3653 | 0.43% | 0.3576 | -1.67% |
| Average share of draws in a season | 0.0226 | 0.2726 | 0.2683 | -1.57% | 0.2847 | 4.44% |
| Number of teams in a season | 0.0551 | 21.774 | 18.451 | -15.26% | 18.039 | -17.16% |

Source: Author's calculations using the data
"V.weight" is the optimal covariate weight in the diagonal of the V matrix.
"Synthetic Control" is the weighted average of control units in the donor pool with optimal weights.
"Average Control" is the simple average of control units in the donor pool with equal weights.

*The effect of the three-point rule on average goals scored per match*

Having obtained the synthetic English league through a linear combination of other leagues, we now compare the outcome in the post-intervention period to measure the average treatment effect of the three-point rule on average goals scored per match in the English league.

As highlighted in Table A8, we find that in 1981-1993, there is a negligible and insignificant decrease in the average number of goals scored per match, as measured by the average treatment effect of -0.0131.

Table A8: Average treatment effect (ATE) and prediction results in post three-point rule period

| Time | Actual Outcome | Predicted Outcome | Treatment Effect |
|---|---|---|---|
| 1981 | 0.9318 | 1.0843 | -0.1526 |
| 1982 | 1.0065 | 1.0739 | -0.0675 |
| 1983 | 0.9953 | 1.0731 | -0.0778 |
| 1984 | 1.0253 | 1.0157 | 0.0096 |
| 1985 | 1.0253 | 1.0236 | 0.0017 |
| 1986 | 0.9669 | 0.9401 | 0.0269 |
| 1987 | 0.9153 | 0.9846 | -0.0692 |
| 1988 | 0.9288 | 0.9617 | -0.0328 |
| 1989 | 0.9535 | 0.9257 | 0.0278 |
| 1990 | 1.0154 | 0.8737 | 0.1418 |
| 1991 | 0.9326 | 0.8926 | 0.04 |
| 1992 | 0.9727 | 0.9722 | 0.0005 |
| 1993 | 0.9503 | 0.9685 | -0.0182 |
| **Mean** | **0.971** | **0.984** | **-0.013** |

Source: Author's calculations using the data

**Appendix A3: Impact of the three-point rule on $\widehat{NAMSI}$ in English league**

*Constructing a synthetic version of the English league*

For the construction of the synthetic version of the English league, we estimate the weights using the following predictors: average share of wins and draws, number of teams count, and all two-period lag values of $\widehat{NAMSI}$ (that is, $t_0, t_2, t_4$ and so on). Table A9 compares treated and synthetic control values for all matching covariates.

Next, we check the weights assigned to the leagues in the donor pool to construct the synthetic English league. We observe that three leagues have positive weights: French 0.3080, Spanish 0.2080 and Dutch 0.4830.

*The effect of the three-point rule on $\widehat{NAMSI}$*

Having obtained the synthetic English league through a linear combination of other leagues, we now compare the outcome in the post-intervention period to measure the average treatment effect of the three-point rule on $\widehat{NAMSI}$ in the English league. As highlighted in table A10, we find that in the period 1981-1993, there is a negligible and insignificant decrease in $\widehat{NAMSI}$, as measured by the average treatment effect of -0.0297.

Table A9: Covariate balance in the pretreatment period

| Covariate | V.weight | Treated | Synthetic value | Control bias | The average value across donor pool | Control bias |
|---|---|---|---|---|---|---|
| NAMSI_HAT(1963) | 0 | 0.3203 | 0.3033 | -5.30% | 0.3672 | 14.64% |
| NAMSI_HAT(1965) | 0 | 0.3264 | 0.4493 | 37.63% | 0.4244 | 30.03% |
| NAMSI_HAT(1967) | 0.4455 | 0.346 | 0.3438 | -0.63% | 0.3478 | 0.52% |
| NAMSI_HAT(1969) | 0 | 0.4072 | 0.4235 | 4.00% | 0.4141 | 1.68% |
| NAMSI_HAT(1971) | 0 | 0.4128 | 0.4715 | 14.20% | 0.4539 | 9.96% |
| NAMSI_HAT(1973) | 0 | 0.2532 | 0.37 | 46.11% | 0.3636 | 43.56% |
| NAMSI_HAT(1975) | 0 | 0.3287 | 0.2921 | -11.14% | 0.3479 | 5.85% |
| NAMSI_HAT(1977) | 0.2751 | 0.3837 | 0.3707 | -3.39% | 0.343 | -10.60% |
| NAMSI_HAT(1979) | 0.0845 | 0.3526 | 0.3675 | 4.24% | 0.3612 | 2.44% |
| Average share of wins in a season | 0.0608 | 0.3637 | 0.3676 | 1.07% | 0.3576 | -1.67% |
| Average share of draws in a season | 0.0298 | 0.2726 | 0.2647 | -2.90% | 0.2847 | 4.44% |
| Number of teams in a season | 0.1043 | 21.7742 | 18.487 | -15.10% | 18.0387 | -17.16% |

Source: Author's calculations using the data
"V.weight" is the optimal covariate weight in the diagonal of the V matrix.
"Synthetic Control" is the weighted average of control units in the donor pool with optimal weights.
"Average Control" is the simple average of control units in the donor pool with equal weights.

| Table A10: Average treatment effect (ATE) and prediction results in post-three-point rule period | | | |
|---|---|---|---|
| Time | Actual Outcome | Predicted Outcome | Treatment Effect |
| 1981 | 0.3679 | 0.4078 | -0.0399 |
| 1982 | 0.29 | 0.41 | -0.12 |
| 1983 | 0.3005 | 0.4241 | -0.1236 |
| 1984 | 0.3695 | 0.3684 | 0.0011 |
| 1985 | 0.4487 | 0.405 | 0.0437 |
| 1986 | 0.331 | 0.3915 | -0.0605 |
| 1987 | 0.3892 | 0.3612 | 0.028 |
| 1988 | 0.3163 | 0.3432 | -0.027 |
| 1989 | 0.3338 | 0.29 | 0.0438 |
| 1990 | 0.3724 | 0.3398 | 0.0327 |
| 1991 | 0.2666 | 0.3237 | -0.0571 |
| 1992 | 0.252 | 0.3752 | -0.1232 |
| 1993 | 0.3805 | 0.3649 | 0.0156 |
| **Mean** | **0.34** | **0.37** | **-0.03** |
| Source: Author's calculations using the data | | | |